\documentclass{pasa}%

\usepackage{graphicx}
\usepackage{epstopdf}
\usepackage{booktabs,caption}
\usepackage[flushleft]{threeparttable}

\title[The most luminous high-redshift QSO]{Discovery of the most ultra-luminous QSO using Gaia, SkyMapper and WISE}

\author[Wolf et al.]{Christian Wolf$^{1,2}$, Fuyan Bian$^{3}$, Christopher A. Onken$^{1,2}$, Brian P. Schmidt$^{1,2}$, Patrick Tisserand$^{1,4}$, Noura Alonzi$^{5,6}$, Wei Jeat Hon$^{5}$ and John L. Tonry$^{7}$
\affil{$^1$Research School of Astronomy and Astrophysics, Australian National University, Canberra, ACT 2611, Australia}
\affil{$^2$ARC Centre of Excellence for All-sky Astrophysics (CAASTRO)}
\affil{$^3$European Southern Observatory (ESO), Vitacura, Chile}
\affil{$^4$Sorbonne Universit\'{e}s, UPMC Univ Paris 6 et CNRS, Institut d`Astrophysique de Paris, 98 bis bd Arago, F-75014 Paris, France}
\affil{$^5$School of Physics, University of Melbourne, Parkville, Victoria 3010, Australia}
\affil{$^6$Department of Physics and Astronomy, King Saud University, Riyadh 11451, Saudi Arabia}
\affil{$^7$Institute for Astronomy, University of Hawaii, 2680 Woodlawn Drive, Honolulu, HI 96822}
}%

\jid{PASA}
\doi{10.1017/pas.\the\year.xxx}
\jyear{\the\year}

\usepackage{aas_macros}
\usepackage{hyperref} 
\hypersetup{colorlinks,citecolor=blue,linkcolor=blue,urlcolor=blue}

\hypersetup{draft}
\begin{document}

\newcommand{\refbf}{} %{} %{\bf }

\begin{frontmatter}
\maketitle

\begin{abstract}
We report the discovery of the ultra-luminous QSO SMSS~J215728.21-360215.1 with magnitude $z=16.9$ and W4$=7.42$ at redshift 4.75. Given absolute magnitudes of $M_{145,\rm AB}=-29.3$, $M_{300,\rm AB}=-30.12$ and $\log L_{\rm bol}/L_{\rm bol,\odot} = 14.84$, it is the QSO with the highest unlensed UV-optical luminosity currently known in the Universe. It was found by combining proper-motion data from {\it Gaia} DR2 with photometry from SkyMapper DR1 and the {\it Wide-field Infrared Survey Explorer} (WISE). In the {\it Gaia} database it is an isolated single source and thus unlikely to be strongly gravitationally lensed. It is also unlikely to be a beamed source as it is not discovered in the radio domain by either NVSS or SUMSS. It is classed as a weak-emission-line QSO and possesses broad absorption line features. A lightcurve from ATLAS spanning the time from October 2015 to December 2017 shows little sign of variability. 
\end{abstract}

\end{frontmatter}

\section{INTRODUCTION}\label{sec:intro}

Black holes at the centres of galaxies reach masses of over ten billion times that of our Sun. Surprisingly we have found such massive black holes already in the early Universe, just 800 million years after the Big Bang \citep{Wu15}. How they grew to such mass so early after the Big Bang is a profound puzzle for physics. They must have grown at super-Eddington rates for a long period of time; or they originate from massive seed black holes that formed during the dark early ages by direct collapse \citep{BrommLoeb03,Pacucci15}.

Currently, {\refbf we can only} discover such super-massive black holes in the distant early universe {\refbf while} they grow rapidly and accrete vast amounts of matter. This makes them appear as very luminous quasi-stellar objects \citep[QSOs,][]{Schmidt63} {\refbf when we have a clear view of the accretion disk around the black hole, and as type-2 QSOs and infrared-luminous galaxies when that view is blocked by dust \citep[e.g.][]{Seymour07,Lacy13}; at present, we can only measure the masses of their black holes when we have a clear view.} 

Finding the most luminous, optically bright, QSOs is important for several reasons: (i) they point us to the most massive black holes that pose the greatest challenge to any physical growth scenario; (ii) they ionise large volumes of neutral gas around them and contribute to cosmic re-ionisation \citep{Fan06a,Wu15}; (iii) they reveal the metal enrichment in the early universe by shining like beacons through the gas content of high-redshift galaxies along the line-of-sight that are otherwise hard to observe \citep{RyanWeber09,Simcoe11}; and (iv) they will eventually enable the most sensitive direct observations of the expansion of the Universe \citep{JoeLis08}.

\begin{figure*}
\begin{center}
\includegraphics[angle=0,width=1.0\textwidth]{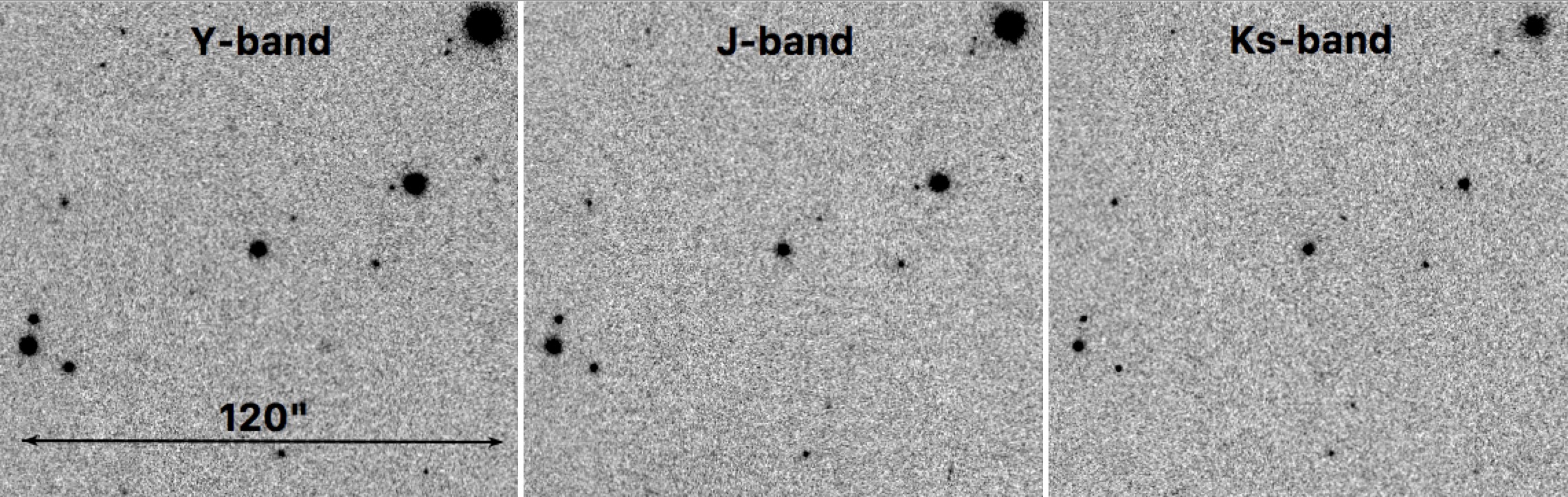}
\caption{Images of J2157-3602 from the VISTA Hemispheric Survey (VHS) in $YJK_{s}$-bands. The QSO appears as an isolated, single point source just like it does in the {\it Gaia} $R_p$-band, so its brightness is unlikely to be boosted by strong gravitational lensing.}
\label{images}
\end{center}
\end{figure*}

Among optically luminous QSOs at high redshift, the most impressive objects are J0306+1853 at $z=5.36$, which is powered by a black hole of 10~billion solar masses \citep{Wang15}, and J0100+2802 at $z=6.3$ with a black hole of 12~billion $M_\odot$ \citep{Wu15}; both of them radiate at their Eddington limit. \citet{Tsai15} claim the bolometrically most luminous object in the Universe is the ELIRG WISE J224607.57-052635.0 at $z=4.593$ with $\log L_{\rm bol} / L_\odot = 14.54$ {\refbf but disagree with \citet{Wu15} on the bolometric correction.}

{\refbf Here, we report a newly identified high-redshift QSO with the highest unlensed UV-optical luminosity} known at present. Throughout the paper, 2MASS and WISE magnitudes are used in the Vega system, while optical magnitudes are in the AB system \citep{Oke83}. We adopt a standard flat $\mathrm{\Lambda CDM}$ cosmology with parameters $H_0=70$~km~s$^{-1}$~Mpc$^{-1}$, $\Omega_{\rm M}=0.3$ and $\Omega_\Lambda= 0.7$.

\section{Data}\label{selection}

\subsection{Selection and photometry}

The discovery of SMSS J215728.21-360215.1 (hereafter J2157-3602) was helped by the Data Release 2 from {\it Gaia} \citep{GaiaDR2} and its unprecedented precision on proper motion measurements. Prior to {\it Gaia} DR2, ultra-luminous QSOs were discovered only serendipitously, because QSO candidate lists were inflated by staggering contamination from cool stars. Even though QSO colours differ from star colours to some degree, the scatter in star colours leads to overlapping loci in colour space. And at bright levels, where the luminosity function of high-z QSOs plummets towards zero space density, stars vastly outnumber real QSOs. Thus, spectroscopic follow-up was often biased against the brightest and potentially most exciting candidates as they were by far the least likely to be QSOs.  

However, the main contaminants of the search box for high-z QSOs are nearby Galactic stars of low mass and temperature, which can now be identified very reliably from the proper motions in the {\it Gaia} database. Hence, immediately after the release of {\it Gaia} DR2 on 25 April 2018, we searched for red objects using $B_p-R_p$ in the {\it Gaia} database that are consistent with having no significant proper motion. Candidates were then cross-matched against SkyMapper DR1 \citep{Wolf18}, the 2 Micron All Sky Survey \citep[2MASS,][]{2MASS} and the {\it Wide-field Infrared Survey Explorer} \citep[WISE,][]{Wright10}, and further selected to have optical-infrared colours appropriate for high-z QSOs (details will be published by Wolf et al., in preparation).

J2157-3602 appears in {\it Gaia} DR2 as a single isolated source; the nearest neighbour detected by {\it Gaia} is $\sim 42$~arcsec away, while the resolution achieved by {\it Gaia} reaches up to $\sim 0.1$~arcsec.\footnote{In \citet{Qi15} the object was listed with an incompatible, large proper motion of $\mu_{\rm dec}= -11.3 +/-1.5$~milliarcsec.} {\it Gaia} is complete to magnitude $R_p\approx 19$, while our object is $\sim 2$~mag brighter; this suggests that the brightness of the object is not boosted by strong gravitational lensing. We find the same conclusion after searching infrared images from the VISTA Hemisphere Survey \citep[VHS,][]{VHS} for signs of an extended source that could signify either multiple images or the presence of a lensing galaxy. In all three bands, $YJK_{s}$, the source is fully consistent with a point source, where the PSF FWHM is 1.04, 0.94 and 0.91 arcsec, respectively (see also Fig.~\ref{images}). {\refbf We checked the \citet{Abell89} catalogue of galaxy clusters in a 2$^\circ$ radius around the QSO and found no entry that could have weakly lensed it.} 

J2157-3602 is clearly detected by SkyMapper in the far-red bands with $i=17.37$ and $z=17.11$, but not in any of the bluer bands, which are only complete to 18~mag in DR1. It was detected by VST-ATLAS \citep{Shanks15} with $g\approx 21$, $r=18.68$, $i=17.32$ and $z=16.93$. Most photometry for high-z QSOs is from the SDSS survey \citep[see e.g.][]{Wang16}, and while VST filters are very similar to the SDSS, the SkyMapper passbands are different. The optical colour is consistent with the $4.7<z<5.4$ QSO selection criteria used by \citet{Wang16}, but it is outside of the optical selection box used by \citet{Richards02}.

\begin{figure*}
\begin{center}
\includegraphics[angle=0,width=0.8\textwidth]{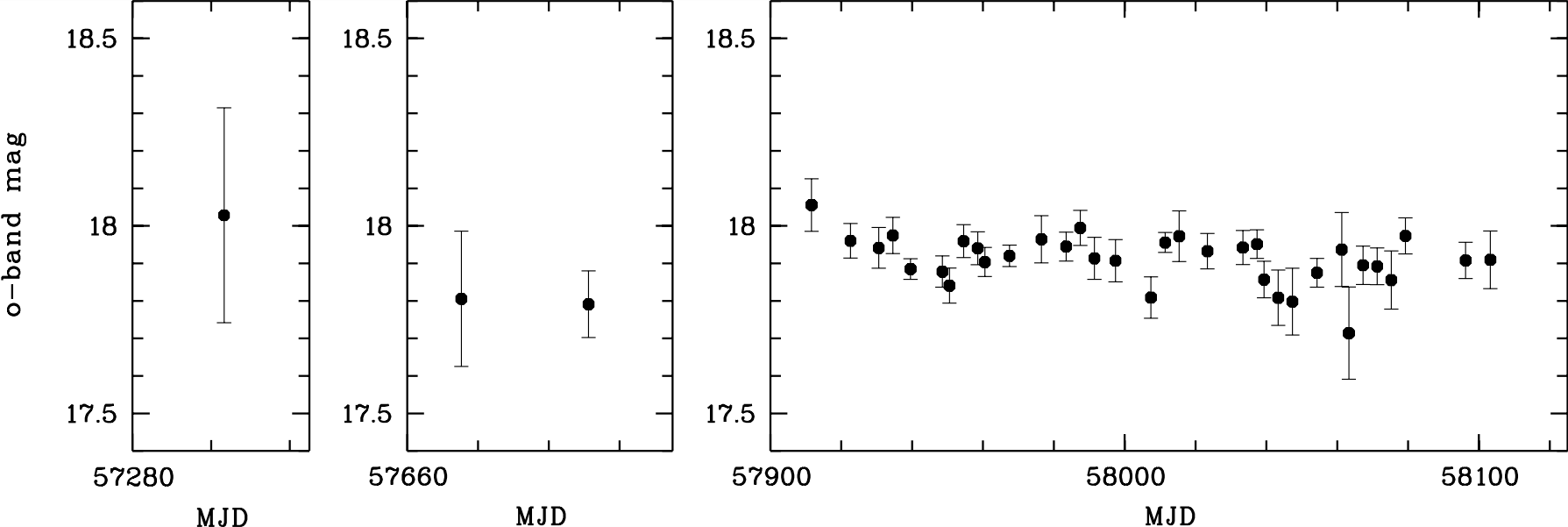}
\caption{Light-curve of J2157-3602 from the Asteroid Terrestrial-impact Last Alert System (ATLAS) project covering three seasons from October 2015 to December 2017, {\refbf after binning magnitudes into 5-day intervals to reduce noise}. There is no sign of strong variability, but slow variations at a $\pm 0.1$~mag level can be seen.}
\label{lightcurve}
\end{center}
\end{figure*}

It was also clearly detected in 2MASS and WISE, reaching magnitude 7.42 in the 22$\mu$-band W4. The WISE position is consistent with the {\it Gaia} position within 0.05~arcsec. The nearest neighbouring source in the WISE catalog is 26~arcsec away and is 4~mag fainter in W1 and 1~mag fainter in W4. With a colour of W1$-$W2$=0.49 \pm 0.02$ the QSO was just outside the selection box used by \citet{Wang16}. Their selection of W1$-$W2$>0.5$ was motivated by suppressing contamination from stars, but with {\it Gaia} it is feasible to push the selection limits further into the stellar locus. 

In Fig.~\ref{lightcurve}, we show the light-curve of the object in the orange filter of the Asteroid Terrestrial-impact Last Alert System \citep[ATLAS,][]{Tonry18} covering three observing seasons from October 2015 to December 2017. The mean error and scatter is $\sim 0.1$~mag and little variability is detected above that, but some variation can be seen after binning the data into 5-day intervals.

J2157-3602 is not detected in any large radio survey: neither by the NRAO-VLA Sky Survey \citep[NVSS,][]{NVSS}, which has a flux limit of 2.5~mJy at 1.4~GHz, nor by the Sydney University Molonglo Southern Survey \citep[SUMSS,][]{SUMSS} with a limit of 5~mJy at 843~MHz. {\refbf We use the radio-to-optical ratio $R=f_{\nu,\rm 6cm}/f_{\nu,B}$ as defined by \citet{Kellermann89} to label the object as radio-loud ($R>10$) or radio-quiet ($R<1$); in that work the ratio is based on the flux density at 6~cm and 440~nm wavelength for a sample with an average redshift of $\sim 0.5$, but for our object these bands are redshifted to $\sim 23$~cm, which is approximately the NVSS wavelength, and 1710~nm. We find a ratio of $R<2.0$ and thus assume that the brightness is not significantly} increased by the presence of a jet with relativistic boosting. Photometric and additional properties of the object are listed in Table~1.

\subsection{Spectroscopy}

We took {\refbf a} spectrum of this object during the full-moon night of 29 April 2018 18:09 UT, using the ANU 2.3m-telescope at Siding Spring Observatory in Australia with the Wide Field Spectrograph \citep[WiFeS,][]{Dopita10}. We {\refbf took two exposures of 600-sec with} the WiFeS R3000 grating that covers the wavelength range from 5300~\AA \ to 9800~\AA \ with a resolution of $R=3000$. The data were reduced using the Python-based pipeline PyWiFeS \citep{Childress14}. The flux density was calibrated using the standard stars HD~55496 and HD~111786 observed on the same night.

\begin{figure}
\begin{center}
\includegraphics[angle=0,width=1.0\columnwidth]{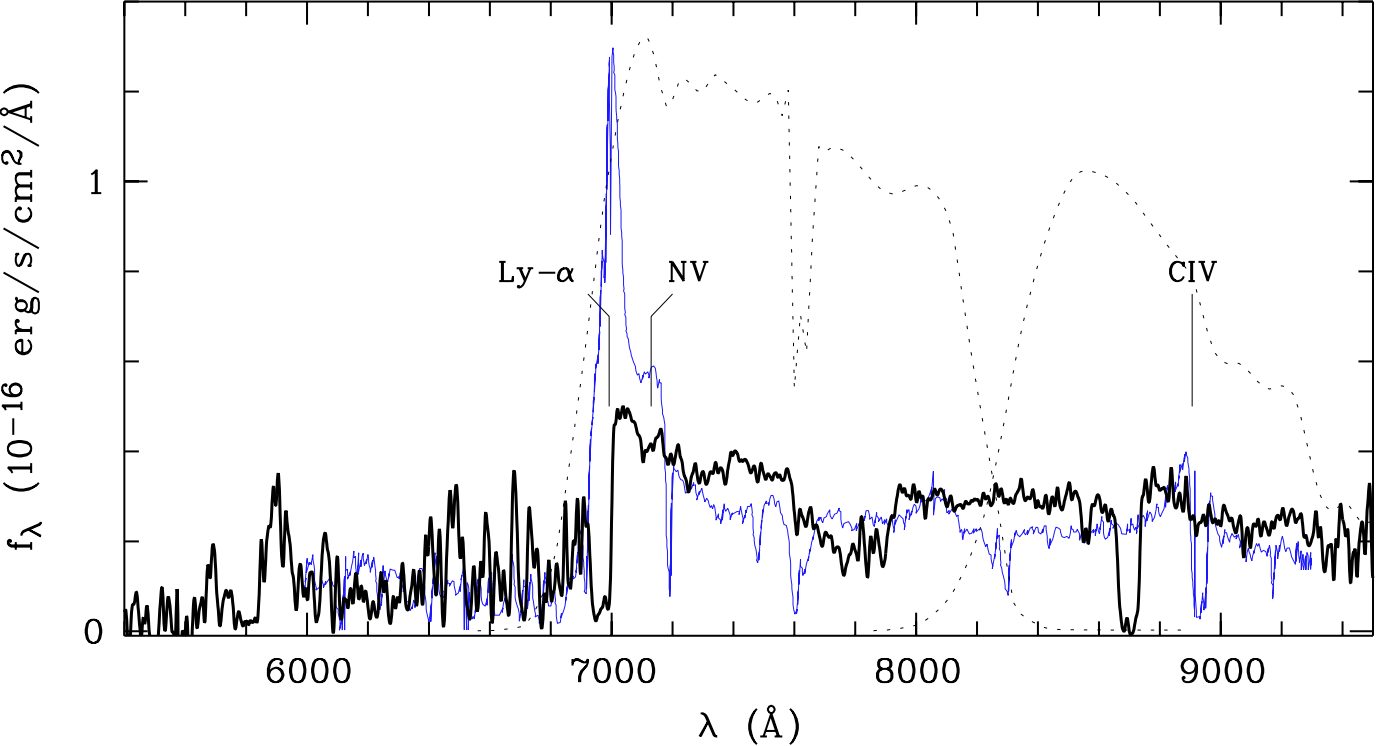}
\caption{Spectrum of J2157-3602 {\refbf (thick line) compared to PMN J1451 (thin line). Strong Ly-$\alpha$ emission makes PMN J1451 brighter in VST $i$-band (dashed) while it is fainter in the continuum and $z$-band. Both spectra are calibrated from photometry.}}
\label{spectrum}
\end{center}
\end{figure}

The spectrum is shown in Fig.~\ref{spectrum} and shows only weak emission lines but a clear Lyman-$\alpha$ forest. The object is thus classified as a Weak-emission-line QSO \citep[WLQ, {\refbf see}][]{Diamond09}. Hints of emission lines are seen for Lyman-$\alpha$ and NV as well as CIV, and based on these we determine the redshift to $z=4.75$. Using Ned Wright's online cosmology calculator\footnote{\url{ http://www.astro.ucla.edu/~wright/CosmoCalc.html}} we find a luminosity distance of $D_L=43.91$~Gpc. Deep broad absorption lines (BALs) are observed on the blue side of the Ly-$\alpha$ and CIV {\refbf line. For} WLQs, a precise redshift determination is hampered by the poorly defined centre of the broad emission lines. {\refbf An alternative redshift of $z=4.64$} is obtained by {\refbf assuming that} the red edge of the {\refbf CIV BAL is} at the systemic velocity of the QSO. 

WLQs are mostly found at high redshift; their fraction among the overall QSO population rises steeply between $z=3$ and $z=4$. The physical explanation for WLQs in not clear, but they are predominantly radio-quiet so their continuum is typically not boosted by radiation from relativistic jets \citep{Diamond09}.

\begin{figure*}
\begin{center}
\includegraphics[angle=0,width=\textwidth]{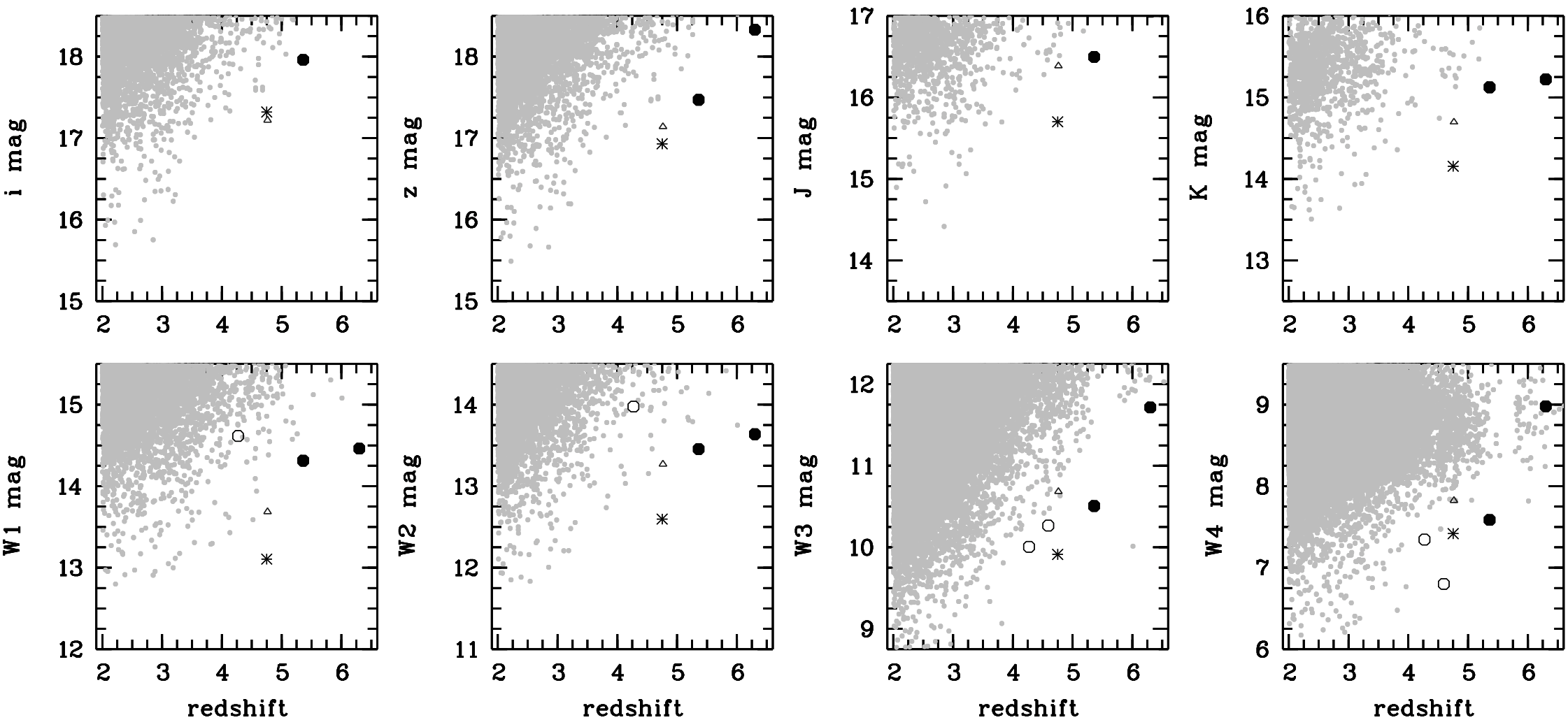}
\caption{Photometry of J2157-3602 (asterisk) at $z=4.75$ compared to known {\refbf QSOs and ELIRGs \citep[grey dots, from][]{Paris17,Wang16,Tsai15}; we highlight PMN J1451-1512 (triangle), the most luminous IR-bright AGN from \citet{Lacy13} at $z=4.27$ and from \citet{Tsai15} at $z=4.59$ (open symbols), and the UV-bright QSOs from \citet{Wang15} at $z=5.36$ and \citet{Wu15} at $z=6.3$ (filled symbols). Magnitudes in $i$- and $z$-bands} are from SDSS and SDSS-like VST.}
\label{photometry}
\end{center}
\end{figure*}

\section{LUMINOSITY VS. OTHER QSOS}\label{comparison}

{\refbf The brightest known UV-bright QSO at $z>4.5$ in the compilation by \citet{Wang16} is J145147.05-151220.20 at $z=4.763$.} Previously labelled PMN J1451-1512, it {\refbf was identified} as a radio-loud QSO by \citet{Hook02} with a 1.4~GHz flux density of $28.5\pm1.0$~mJy in NVSS. The object is not contained in the SDSS footprint, but in the SDSS-like filters of VST-ATLAS it has $i=17.22$ and $z=17.14$. It is not listed in the SkyMapper DR1 'master' table as all its data have warning flags due to proximity to a bright star (the V$\sim$7.8~mag HD~131046, 100~arcsec away); however, visual inspection reveals that the images are clean and we retrieve $i=17.23\pm 0.05$ and $z=17.19\pm 0.1$ from the 'fs\_photometry' table that contains all detections. 

So, PMN J1451-1512 is brighter by $\sim 0.1$~mag than our new object in both the $i$-band and in the {\it Gaia} $R_p$-band, but $\sim 0.1$~mag fainter in the $z$-band. This is related to the lack of strong emission lines and the presence of broad absorption bands in our object. PMN J1451-1512 has strong Ly-$\alpha$ emission with $\sim 40$~nm equivalent width (see Fig.~\ref{spectrum}), which boosts its $i$-magnitude by $\sim 0.3$~mag, while in the continuum between the Ly-$\alpha$ and CIV lines our object is 0.2~mag brighter than PMN J1451-1512.

\begin{table}
\label{Table_properties}
\begin{center}
\caption{Properties of QSO J2157-3602.}
\begin{tabular}{lr}
\hline
\multicolumn{2}{|c|}{Gaia positional data}   \\
\hline
RA (deg)		& $329.3676229$ \\
Dec (deg)		& $-36.0375573$ \\
$b_{\rm gal}$	& $-52.2195997$ \\
$l_{\rm gal}$	& $  8.2221573$ \\
parallax (milli-arcsec) & $0.15 \pm 0.33$ \\
proper motion (milli-arcsec/yr) & $0.03 \pm 0.56$ \\
ref-epoch	& 2015.5 \\
$E(B-V)_{\rm SFD}$ & 0.0147 \\
\hline
\multicolumn{2}{|c|}{Photometry} \\
\hline
Gaia $B_p$	& $19.794 \pm 0.058$ \\
Gaia $G$	& $18.286 \pm 0.003$ \\
Gaia $R_p$	& $17.136 \pm 0.007$ \\
VST-ATLAS $u_{\rm AB}$ & undetected \\
VST-ATLAS $g_{\rm AB}$ & $20.94 \pm 0.03$\\
VST-ATLAS $r_{\rm AB}$ & $18.68 \pm 0.01$ \\
VST-ATLAS $i_{\rm AB}$ & $17.32 \pm 0.01$ \\
VST-ATLAS $z_{\rm AB}$ & $16.93 \pm 0.01$ \\
SkyMapper $r_{\rm AB}$  & undetected \\
SkyMapper $i_{\rm AB}$  & $17.37 \pm 0.02$ \\
SkyMapper $z_{\rm AB}$  & $17.11 \pm 0.02$ \\
VHS $Y$ 	& $16.15 \pm 0.01$ \\
VHS $J$ 	& $15.65 \pm 0.01$ \\
VHS $K_{s}$ & $14.25 \pm 0.01$ \\
2MASS $J$	& $15.70 \pm 0.06$ \\
2MASS $H$	& $14.84 \pm 0.05$ \\
2MASS $K$	& $14.16 \pm 0.06$ \\
WISE W1	{\refbf (mpro)} & {\refbf $13.11 \pm 0.02$} \\
WISE W2	{\refbf (mpro)} & {\refbf $12.60 \pm 0.03$} \\
WISE W3	{\refbf (mpro)} & {\refbf $ 9.91 \pm 0.05$} \\
WISE W4	{\refbf (mpro)} & {\refbf $ 7.42 \pm 0.13$} \\
NVSS $f_{\rm 1.4GHz}$ & $<2.5$~mJy \\
SUMSS $f_{\rm 843MHz}$ & $<5.0$~mJy \\
\hline
\multicolumn{2}{|c|}{Redshift and luminosities} \\
\hline
Redshift & $\sim 4.75$ \\
$t_{\rm look-back}$ & 12.23~Gyr \\
$M_{145,\rm AB}$ & $-29.30$ \\
$M_{300,\rm AB}$ & $-30.12$ \\
$M_{\rm bol}$ & $-32.36$ \\
$L_{\rm bol}/L_\odot$ & $6.95\times 10^{14}$ \\
\hline
\end{tabular}
\end{center}
\end{table}

Our object is consistently brighter than PMN J1451-1512 by $0.5\pm 0.05$~mag across the whole infrared range from 1.2$\mu$ ($J$-band) to 22$\mu$ (W4), suggesting a nearly identical spectral slope across the rest-frame range from 200~nm to 4$\mu$, and thus will have a larger bolometric energy output (see Figure~\ref{photometry}). On the blue side of 200~nm rest-frame, the continuum appears redder than typical for QSOs, which might have the same origin as the broad absorptions, but we leave the discussion of this point until we obtain a detailed high-S/N spectrum.

We estimate continuum luminosities at rest-frame 145 and 300~nm from broadband photometry in $z$- and $HK$-bands, respectively, because our early spectrum is unfortunately too noisy and its calibration too uncertain to derive it from that; due to the weak emission and broad absorption lines, we assume this to be an appropriate estimate. We find $M_{145,\rm AB}=-29.30$ and $M_{300,\rm AB}=-30.12$. We translate this into monochromatic luminosities of $\nu L_{\nu,145}=4.71 \times 10^{47}$~ergs~s$^{-1}$ and $\nu L_{\nu,300}=4.75 \times 10^{47}$~ergs~s$^{-1}$. Using bolometric corrections of 3.8 and 5.6 as estimated in \citet{Richards06}, we find two values for the bolometric luminosity of $L_{\rm bol}=1.8 \times 10^{41}$~W and $2.66 \times 10^{41}$~W. We adopt the latter value as all wavelengths above 200~nm rest-frame support it; it corresponds to $L_{\rm bol}/L_\odot = 6.95\times 10^{14}$ and an absolute bolometric magnitude $M_{\rm bol}=-32.36$. 

Thus, our new object would be more luminous than QSO J0100+2802 and ELIRG J2246-0526, although bolometric {\refbf corrections remain uncertain. However, as Fig.~\ref{photometry} shows, the monochromatic luminosity of ELIRG J2246-0526 exceeds that of J2157-3602 at restframe 4$\mu$ (W4-band) and most probably longwards of that. We note, that the lensed QSO APM 08279+5255 at $z=3.9$ has a magnified luminosity exceeding that of our object, but a lower intrinsic luminosity \citep{Ibata99}. }

Lacking a clearly rendered CIV line, {\refbf we cannot} get a reliable estimate of the black-hole mass at the moment. For the two most luminous $z>5$ QSOs black-hole masses were estimated from their MgII lines: J0306+1853 at $z=5.36$ is powered by a black hole of 10~billion solar masses \citep{Wang15} and J0100+2802 at $z=6.3$ has a black hole of 12~billion solar masses \citep{Wu15}. Both {\refbf of} these objects appear to radiate at the Eddington limit, and if J2157-3602 {\refbf does the same}, then we expect its black hole to have 20 billion solar masses.

\section{SUMMARY}\label{summary}

ESA's {\it Gaia} mission, with its unprecedented precision of proper motion measurements, has allowed us to overcome the main obstacle for the discovery of ultra-luminous QSOs in the early universe, which is the staggering contamination of candidate lists by cool stars in our own Galaxy. By combining data from Gaia, SkyMapper and WISE, we identified SMSS~J215728.21-360215.1 as the brightest $z\sim 5$ QSO known at this point; it has a magnitude of $z_{\rm VST}=16.93$ and $H_{\rm 2MASS}=14.84$. With a redshift of $z=4.75$ this translates into an absolute magnitude of $M_{300,\rm AB}=-30.12$ and a bolometric luminosity of $6.95\times 10^{14}$ solar luminosities.

It is not detected in the radio surveys NVSS and SUMSS and thus not likely to be boosted in optical brightness by a relativistic jet. It appears as a single isolated point source in all images and catalogues we considered, most notably the {\it Gaia} DR2 database and infrared images from the VISTA Hemisphere Survey, and thus not likely to be strongly gravitationally lensed. The derived luminosity is {\refbf likely intrinsic}, making J2157-3602 the {\refbf QSO with the highest unlensed UV-optical luminosity} discovered so far. Our current estimate for the mass of its black hole is 20 billion solar masses, assuming it accretes at the Eddington limit.

While objects of this luminosity are exceedingly rare in the Universe, they are particularly valuable as bright background and reference sources in order to study the properties of intervening matter along the line-of-sight, and for directly probing the expansion of our Universe with new instruments in the coming decades. 

Finding high-redshift QSOs is even more efficient when the follow-up of candidates is embedded in a large spectroscopic survey. Later this year, the hemispheric Taipan survey will commence at Siding Spring Observatory and take spectra of more than a million galaxies \citep{daCunha17}. A program that piggybacks on Taipan aims to go after the $10^4$ best candidates and will confirm a complete sample of supermassive black holes with luminous accretion in the early Universe at $4<z<7$.

\begin{acknowledgements}
This research was conducted by the Australian Research Council Centre of Excellence for All-sky Astrophysics (CAASTRO), through project number CE110001020.
It has made use of data from the European Space Agency (ESA) mission {\it Gaia} (\url{https://www.cosmos.esa.int/gaia}), processed by the {\it Gaia} Data Processing and Analysis Consortium (DPAC, \url{https://www.cosmos.esa.int/web/gaia/dpac/consortium}). Funding for the DPAC has been provided by national institutions, in particular the institutions participating in the {\it Gaia} Multilateral Agreement.
The national facility capability for SkyMapper has been funded through ARC LIEF grant LE130100104 from the Australian Research Council, awarded to the University of Sydney, the Australian National University, Swinburne University of Technology, the University of Queensland, the University of Western Australia, the University of Melbourne, Curtin University of Technology, Monash University and the Australian Astronomical Observatory. SkyMapper is owned and operated by The Australian National University's Research School of Astronomy and Astrophysics. The survey data were processed and provided by the SkyMapper Team at ANU. The SkyMapper node of the All-Sky Virtual Observatory (ASVO) is hosted at the National Computational Infrastructure (NCI). Development and support the SkyMapper node of the ASVO has been funded in part by Astronomy Australia Limited (AAL) and the Australian Government through the Commonwealth's Education Investment Fund (EIF) and National Collaborative Research Infrastructure Strategy (NCRIS), particularly the National eResearch Collaboration Tools and Resources (NeCTAR) and the Australian National Data Service Projects (ANDS).
This work uses data products from the Wide-field Infrared Survey Explorer, which is a joint project of the University of California, Los Angeles, and the Jet Propulsion Laboratory/California Institute of Technology, funded by the National Aeronautics and Space Administration.
It uses data products from the Two Micron All Sky Survey, which is a joint project of the University of Massachusetts and the Infrared Processing and Analysis Center/California Institute of Technology, funded by the National Aeronautics and Space Administration and the National Science Foundation.
This paper uses data from the VISTA Hemisphere Survey ESO programme ID: 179.A-2010 (PI. McMahon).
Support for this work was provided by NASA grant NN12AR55G.
\end{acknowledgements}

\bibliographystyle{pasa-mnras}

\iffalse

{\bf ChrisW must check M145 against one object from COMBO-17 in S11 field at z 4.75 with flux of 0.08 photons/m2/nm/sec, but careful COMBO-17 M145 values are Vega, where Vega has 0.42e8 photons/m2/nm/sec and H0 = 65. WHAT IS THE M145 mag of that object in the COMBO-17 table? OUR QSO here has 10.55 photons/m2/nm/sec, so is 5.25 mag brighter.}
\fi 

\end{document}